\documentclass[preprint,showpacs,preprintnumbers,amsmath,amssymb]{revtex4}
 
\usepackage[dvips]{graphicx}
\usepackage{dcolumn}
\usepackage{bm}
\usepackage{amssymb}
 
\begin{document}

\title{ Nonlinear Response With Dichotomous Noise}

\author{I. Bena and C. Van den Broeck}
\affiliation{Limburgs Universitair Centrum, B-3590 Diepenbeek, Belgium}

\author{R. Kawai}
\affiliation{Department of Physics, University of Alabama at Birmingham,
Birmingham,  AL 35294}

\author{Katja Lindenberg}
\affiliation{Department of Chemistry and Biochemistry 0340 and Institute
for Nonlinear Science, University of California, San Diego,
La Jolla, CA 92093}

\date{\today}

\begin{abstract}
Dichotomous noise appears in a wide variety of physical and mathematical
models.  It has escaped attention that the standard results for
the long time properties cannot be applied when unstable fixed points are
crossed in the asymptotic regime. We show
how calculations have to be modified to deal with these cases
and present as a first application
full analytic results for  hypersensitive transport.
\end{abstract}


\maketitle

While the Wiener process together with
its ``time derivative," the Gaussian white
noise, is certainly the method of choice to describe Brownian motion, the
motion induced by another fundamental stochastic process, namely, the
dichotomous Markov process (see, e.g.~\cite{kampen,fulinsky94}),
has its own virtues and interest.
Systems driven
by dichotomous noise can often be described analytically, including
the Gaussian white noise case as a
specific limit, and allow the analytic investigation of the effects of the
finite correlation time of the noise, notably in 
noise induced transitions, noise
induced phase transitions, stochastic resonance, and ratchets. 
Dichotomous noise is often a
good representation of the actual physical
situation, e.g., thermal transitions between two
configurations or states, and
can easily be implemented as an external noise,
with the additional advantage that the support of this noise is finite.

A widely studied generic stochastic equation that describes the
temporal evolution of a single scalar variable $x(t)$ is
\begin{equation}
\dot{x}=\xi(t)v(x)+F,
\label{1}
\end{equation}
where the dot stands for the time derivative, $\xi(t)$ is a symmetric
dichotomous Markov process that takes on the values $\pm 1$ with transition
rate $k$, $v(x)$ is a given velocity profile,
and $F$ is a constant external force. One can of course embellish
this description
in a variety of ways such as, e.g., by allowing for a state- and/or
time-dependent external
force, but here we adhere to this simple
form.  Existing results include the steady state
distribution~\cite{klyatskin77,horsthemke} and first passage time
moments, see e.g.~\cite{MFPT}.
When Eq.~(\ref{1}) is defined on an interval with periodic
boundary conditions, one can evaluate the stationary probability
flux and from it
the average asymptotic drift velocity or diffusion coefficient,
a problem that has recently received a great
deal of attention in the context of Josephson junctions and
Brownian motors~\cite{stokesdrift}.  Although these results 
are often claimed to be completely general, our study shows that
this is {\em not} the case.  Indeed, to our knowledge,  with the
exception of~\cite{bala01},
all the existing results are limited to motion that asymptotically
{\em does not cross unstable fixed points of the dynamics}.  
The main purpose
of our work is to point out where the existing results break down, and to
present the procedure to obtain a fully general solution for the
asymptotic average velocity, including, as a first direct application, the
problem of hypersensitive response~\cite{ginzburg}.

\begin{figure}[htb]
\begin{center}
\centering\includegraphics[width=2.8in]{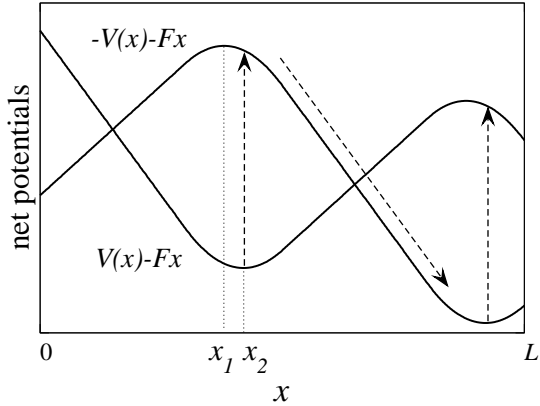}
\end{center}
\caption{
The net potentials $\mp V(x) -Fx$.
}
\label{fig1}
\end{figure}

Consider, then, the prototypical stochastic differential equation
(\ref{1}).  We take the velocity profile $v(x)$ to be
periodic,
$v(x)=v(x+L)$, with zero average, $\int_{0}^{L} v(x)\,dx =0$.
A schematic representation of the two configurations assumed by
the ``net
potentials" $\mp V(x) -Fx$ associated with right hand side (with $v(x)\equiv 
-dV(x)/dx$) is shown in Fig.~\ref{fig1}.  The fixed points
of the dynamics are the points
at which $\pm v(x)+F$ vanish, i.e., the local extrema of the net
potentials.
The stochastic dynamics (\ref{1}) can equivalently be
described by the
master equation for the probability densities $P_{+}(x,t)$ and $P_{-}(x,t)$
for being at $x$ at time $t$ if $\xi=+1$ and $-1$, respectively, with
$x \in [0,L]$ and periodic boundary conditions:
\begin{eqnarray}
\frac{\partial P_{\pm}(x,t)}{\partial
t}&=&\mp\frac{\partial}{\partial x}
\left\{\left[v(x)\pm F\right]P_{\pm}(x,t)\right\}
\nonumber\\
&&-k\left[P_{\pm}(x,t)-
P_{\mp}(x,t)\right].
\label{4}
\end{eqnarray}

To find analytic expressions for the asymptotic steady state
(i.e., time-independent) probabilities and
mean velocity $\left<\dot{x}\right>$, it is more convenient
to introduce the sum and difference probability densities
$P(x,t)=P_{+}(x,t)+P_{-}(x,t)$ and $p(x,t)=P_{+}(x,t)-P_{-}(x,t)$.
Summation of the equations in~(\ref{4}) then immediately leads to the
conclusion that in the steady state the
probability flux $J$ associated with $P(x)$, namely,
$J=FP(x)+v(x)p(x)$,
is a constant whose value is to be determined.  It also leads to
a direct relation between the mean velocity in the stationary state and the
flux,
$\langle \dot{x} \rangle = \int_0^L \left\{ \left[ v(x)+F\right] P_+(x) +
\left[ -v(x)+F\right]P_-(x)\right\} dx = LJ$.

Again in the steady state, subtraction of the
equations in (\ref{4}) and the constant flux condition leads to
the following first-order differential equation for $p(x)$:
\begin{equation}
\left[F^2-v^2(x)\right]
\frac{dp(x)}{dx}-2\left[v(x)\,v'(x)-kF\right]p(x)+Jv'(x)=0,
\label{15}
\end{equation}
where $v'(x)=dv(x)/dx$.
The solution to this equation, together with the constant flux condition
and
the normalization condition
for $P(x)$, $\int_0^LP(x)dx=1$, 
can be used to determine $P(x)$, the flux $J$ and the mean velocity $\left<
\dot{x}\right>$.

The crux of the problem now resides in finding the solution to
Eq.~(\ref{15}). This solution is straightforward 
when $\left[F^2-v^2(x)\right]$
{\em has no zeroes}, that is, when
the net potentials have no extrema within the interval $(0,L)$.
In this case the standard method of variation of parameters leads to the
familiar solution
\begin{equation}
p(x)=-\frac{J}{\left|F^2-v^2(x)\right|}\left[ CG(x,0)+ H(x,0;x)\right],
\label{17}
\end{equation}
where $C$ is a constant of integration that arises from the general solution
to the homogeneous part of Eq.~(\ref{15}), the second contribution
is the particular solution, 
and we have defined the functions
\begin{eqnarray}
H(z,y;x) &=&\int_y^z\mbox{sgn}[F^2-v^2(x')]v'(x') G(x,x')dx',
\nonumber\\
G(z,y)&\equiv& \exp\left[-2kF\int_y^z \frac{dx}{F^2-v^2(x)}\right].
\label{new}
\end{eqnarray}
The usual procedure to determine $C$ is to require periodicity of
$p(x)$, recalling that $v(x)$ is periodic.
One finally obtains $p(x)=[J-FP(x)]/Jv(x)$, and
\begin{eqnarray}
P(x)&&= \frac{J}{F} \left\{1-\frac{v(x)}{\left[1-G(0,L)\right]
[F^2-v^2(x)]}\right. \nonumber\\
&&\left. \times \int_x^{x+L}dx' v'(x')G(x,x')  \right\}.
\label{21}
\end{eqnarray}
The normalization of $P(x)$ determines the value of the flux $J$ and leads
to the following
result for the mean velocity at the steady state:
\begin{eqnarray}
\langle \dot{x}\rangle &=& F \left\{1-\frac{1}{L\left[ 1-G(0,L)\right]}
\int_0^L dx \frac{v(x)}{F^2-v^2(x)} \right. \nonumber\\
&&\left. \times \int_x^{x+L} dx' v'(x') G(x,x') 
\right\}^{-1}.
\end{eqnarray}

The standard results shown above are applicable not only in the
absence of fixed points, but also when the asymptotic behavior
is governed by {\em stable} fixed points.  In this latter case
the dynamics  settles into an alternating motion between these
points, so that they delimit the interval in which the steady state
probability is non-zero~\cite{horsthemke}.  The associated
normalizable divergences
at the fixed points represent regions where the probability density
for finding the system is high.

The situation is entirely different, both physically and
mathematically, 
when the system can cross {\it unstable} fixed points within the interval
$(0,L)$ in the long time limit. A
simple illustration is provided by the example $v(x)=\sin\,x$. In the absence
of an
external force, the dynamics is restricted to an interval
$[k\pi,\,(k+1)\pi]\,\,\,(k\,\,\,{\mbox {integer}})$. Even
though the application of an external forcing
$|F|<1$ cannot induce  ``escape" from this interval
in either of the separate dynamics
$\dot{x}=\sin\, x + F$ and $\dot{x}=-\sin\, x + F$,
running solutions with finite average velocity
appear when the dynamics switches back and forth between the
two~\cite{ginzburg} (see Fig.~\ref{fig1}). The
explicit calculation of this velocity is one of our main goals.
Clearly, the solution~(\ref{21}) is no longer 
correct because it contains non-integrable singularities (see below) 
at the unstable fixed
points where the probability of finding
the system is expected to be {\em low}, not high.

For simplicity we restrict our presentation
to velocity profiles $v(x)$ that are continuously decreasing functions of
$x$ on $[0,\,L/2]$ and symmetric about $L/2$, $\,v(x+L/2)=-v(x)$.
This implies that $P(x+L/2)=P(x)$ and $p(x+L/2)=-p(x)$,
so that we can  limit our analysis to the
interval $[0,\,L/2]$. In this ``minimal scenario",
the equation $F^2\,-\,v^2(x)=0$ has only two solutions in the interval
$[0,\,L/2]$, namely, $x_1$, corresponding to
an unstable fixed point in the
$\xi=-1$ dynamics $[F=v(x_1)]$, and $x_2$, a stable 
fixed point in the $\xi=+1$ dynamics $[F=-v(x_2)]$, with $x_2>x_1$.
The steady state
results leading to Eq.~(\ref{15}) still apply, but the solution to
Eq.~(\ref{15}) is more delicate than the ``blind" integration that
yields Eq.~(\ref{21}). Indeed, the  coefficient of the first derivative is
zero at the fixed points, which now lie
entirely within the support of the probability distribution.  Thus, the
equation becomes singular.

The method of variation of parameters for an equation of the type
(\ref{15}) leads to a solution which is a sum of the general solution
of the homogeneous equation and a particular solution of the
inhomogeneous equation, as in Eq.~(\ref{17}).
The subtlety lies in the determination of the constant of integration $C$,
which in the previous case
was fixed simply by imposing periodic boundary conditions.
In the vicinity of the stable fixed point $x_2$ this straightforward
procedure leads to
the dependence
$P(x)\sim {|x-x_2|^{{k}/{|v'(x_2)|}-1}}$, which is  continuous
when ${k}/{|v'(x_2)|} >1$ and divergent but integrable for
${k}/{|v'(x_2)|}\leq 1$. This result causes no mathematical difficulty and
is consistent with the physical intuition that probability near
a stable fixed point should indeed build up, especially
when the switching rate is low. At
the unstable fixed point $x_1$, however, this procedure leads to an
apparent non-integrable divergence,
$P(x) \sim {|x-x_1|^{-{k}/{|v'(x_1)|}-1}}$,
which is clearly unphysical and
mathematically improper in view of the requirement of normalization. The
fallacy lies in the assumption that a single constant $C$ is valid
throughout the region $(0,L)$.  In fact,
the solution (\ref{17}) is valid in the separate intervals
$[0,x_1)$, $(x_1,x_2)$, and $(x_2,L/2]$, {\em but not necessarily with
the same constant of integration} in all of them.  Indeed, there
is {\em exactly one}
choice of this constant valid for both $[0,x_1)$ and $(x_1,x_2)$ such
that the divergence at $x_1$ is removed, and another choice valid in
the interval $(x_2,L/2]$ that ensures required continuity and periodicity.
In other words, even
though the general solution of the homogeneous equation always has a
divergence, there exists
exactly one solution to the full inhomogeneous equation (\ref{15}) that has
no divergence and is actually completely smooth at
$x_1$. This solution is given by Eq.~(\ref{17}) with the choice
$C=- H(x_1,0;0)$
in both intervals $[0,x_1)$ and $(x_1,x_2)$~\cite{elsewhere}. 
This choice insures that the
coefficient of the divergent term vanishes at $x=x_1$.
We conclude that for ${ x \in [0,x_2)}$,
\begin{equation}
P(x)=\frac{J}{F}\left[1+\frac{v(x)}{\left|F^2-v^2(x)\right|}
H(x,x_1;x)\right].
\label{28}
\end{equation}
Note that $P(x)$ is now continuous at $x_1$, and that 
$\lim_{x \searrow x_1}P(x)=\lim_{x \nearrow
x_1}P(x)={J}F^{-1}\left\{1-[2(k/|v'(x_1)|+1)]^{-1}\right\}$
is indeed finite.

For $x\in (x_2,L/2]$, the result
(\ref{17}) for  $p(x)$  applies again, but now the constant
$C$ is determined by imposing the continuity of $p(x)$ at $x=L/2$.
One finds:
\begin{eqnarray}
P(x)&=&\frac{J}{F}\left\{1+\frac{v(x)}
{\left|F^2-v^2(x)\right|} \left[
H(x,L/2;x) \right. \right. \nonumber\\
&& \left. \left. +G(0,L/2)H(x_1,0;x)\right]\right\}.
\label{32}
\end{eqnarray}
At the stable fixed point $x_2$, $P(x)$ has the behavior described earlier,
i.e., it is continuous for $k/|v_2'(x_2)|>1$ and divergent but integrable
for $k/|v'(x_2)| \leq 1$.

The values of the flux $J$ and of the average velocity follow from these
results by imposing the normalization of $P(x)$: 
\begin{eqnarray}
\langle \dot{x}\rangle &=& F\left\{ 1+\frac{2}{L} \int_0^{x_2} dx
\frac{v(x)}{\left|F^2-v^2(x)\right|} H(x,x_1;x)\right. 
\nonumber\\
&&\left. + \frac{2}{L}\int_{x_2}^{L/2} dx
\frac{v(x)}{\left|F^2-v^2(x)\right|} \left[
H(x,L/2;x) \right. \right. \nonumber\\
&& \left. \left. +G(0,L/2)H(x,0;x)\right]\right\}^{-1}.
\label{eq34}
\end{eqnarray}
This is our main new result. Note that the above procedure can be
repeated straightforwardly but tediously for more complicated
cases involving several stable and unstable fixed points.

\begin{figure}[htb]
\begin{center}
\centering\includegraphics[width=2.8in]{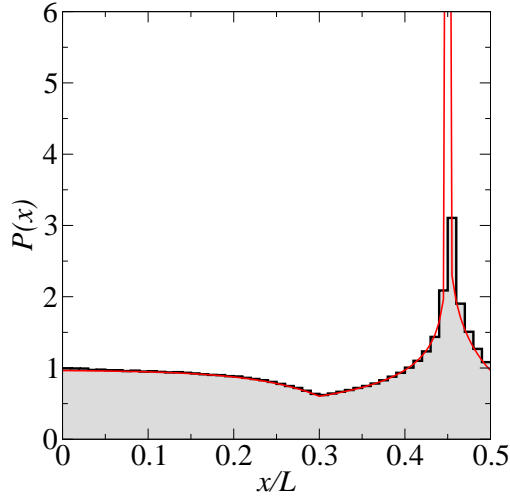}
\end{center}
\caption{
Probability density $P(x)$ vs $x/L$ for the parameter values $f=0.5$,
$\Gamma=0.4$, and $\alpha=1.0$. Histogram: simulation results. Curve: exact
theory.
}
\label{fig2}
\end{figure}

To illustrate our findings with explicit results,
we turn to a particular case of a piecewise linear profile:
\begin{equation}
v(x)=\left\{
\begin{array}{ll}
v_0, &\mbox{for}\,\,\, x\,\in [0,\,L/2\,-\,2\,l) \\
v_0\,\left(\displaystyle \frac{L}{2\,l}\,-\,1\,-\,
\displaystyle \frac{x}{l}\right),
&
\mbox{for}\,\,\,x\,\in\,[L/2\,-\,2\,l,\,L/2) \\ -v(x\,-L/2)\,\,,
&\mbox{for}\,\,\, x\,
\in [L/2,\,L)
\end{array}
\right.
\label{eq35}
\end{equation}
with $l\leq L/4$ and, of course, the periodicity condition
$v(x+L)=v(x)$.  It is convenient to introduce the following
dimensionless variables:
\begin{equation}
f=F/v_0 , \qquad \alpha=lk/v_0 , \qquad \Gamma=4l/L.
\label{36}
\end{equation}
\begin{figure}[htb]
\begin{center}
\centering\includegraphics[width=2.8in]{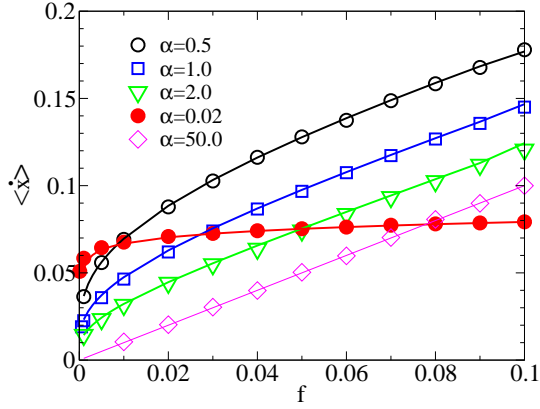}
\end{center}
\caption{
Mean velocity as a function of the applied force for various values of
$\alpha$.  Note the hypersensitive response. 
}
\label{fig3}
\end{figure}
In this case the function $H(x,0;0)/v_0\equiv T(x)$
becomes $T(x)=0$ for $x\in [0,L/2-2l)$ and, for $x\in [L/2-2l, L/2)$,
\begin{eqnarray}
T(x)&=&
\left|\frac{1+f}{1-f}\right|^{\alpha}\exp\left[-\frac{4\alpha f(1-\Gamma)}
{(1-f^2)\Gamma}\right] \nonumber\\
&\times& \int_1^{L/2l-1-x/l}\mbox{sgn}(f^2-s^2)
\left|\frac{f-s}{f+s}\right|^{\alpha} ds.
\label{eq38}
\end{eqnarray}
Explicit {\em exact} results for the probability densities and for
the resultant
average velocity for all values of $f$ are available
and will be detailed elsewhere~\cite{elsewhere}.
Here we exhibit only some of these
results for the new case $0<f<1$.
Figure~\ref{fig2} depicts a typical probability density $P(x)$ vs $x/L$
that clearly shows the agreement between the exact theoretical results and
simulations.  For the average velocity we find
\begin{eqnarray}
\langle \dot{x} \rangle &=&
F\left\{1-\frac{\Gamma T(x_1)}{4\alpha f}
\left[\exp\left(\frac{4\alpha f(1-\Gamma)}{\Gamma(1-f^2)}\right)
-1\right] \right.
\nonumber\\
&&\hspace{-1cm}+\,\frac{\Gamma(1-f)^{\alpha}}{2(1+f)^{\alpha}}
\exp\left(\frac{4\alpha f(1-\Gamma)}{\Gamma(1-f^2)}\right)\times
\nonumber\\
&&\hspace{-1cm}\times\left\{ \int_{-1}^{1}dt\,
\frac{t\,\left[T[L/2-l(t+1)]-T(x_1) \right]}
{\left|t-f\right|^{1+\alpha}\left|t+f\right|^{1-\alpha}}\right.
\nonumber\\
&&\hspace{-1cm}+ \left[T(x_1)
\left(1+ \displaystyle\left(\frac{1+f}{1-f}\right)^{2\alpha}
\exp\left(-\frac{4\alpha f(1-\Gamma)}{\Gamma(1-f^2)}\right)
\right)\right.\nonumber\\
&&\left.\left.\left.\hspace{-1cm}-T(L/2)\displaystyle\right]
\int_{-1}^{-f}dt\, \frac{t}
{\left|t-f\right|^{1+\alpha}\left|t+f\right|^{1-\alpha}}
\right\}\right\}^{-1 }.
\label{velocity}
\end{eqnarray}

The above integrals can be
evaluated explicitly for specific values of $\alpha$,
in particular, for $\alpha=1/2$, $1$, and $2$~\cite{elsewhere}. 
The analytic and simulation results for the variation of the average
velocity with $f$ (again with full agreement) are shown in Fig.~\ref{fig3}.
In the limit of slow switching rate, that is, in the adiabatic regime
$\alpha \rightarrow 0$, one recovers the region of hypersensitive response
discussed in~\cite{ginzburg}, namely,
$\langle\dot{x}\rangle\approx 2v_0\alpha/\Gamma$.
The physics of this result is explained as follows: 
The stable and unstable fixed points of the dynamics, which coincide in the
absence of forcing, are shifted apart by an amount of order of
$ l F /v_0$ for a small force.
Upon each switch of the dichotomous process and for sufficiently slow
switching rate, the particle will glide down to the next stable fixed point,
crossing the location of the unstable fixed point of the alternate
dynamics, cf. Fig.~\ref{fig1}.
As the average time between switches is $k^{-1}$, and the distance
covered is $L/2$, the mean velocity is just $\langle \dot{x} \rangle =
Lk/2$, independently of $F$ and $v_0$.
The typical time $\tau$
for a particle to escape the region around the fixed point is determined
by the relation  $(lF/v_0)\exp(v_0 \tau/l) \approx l$. The crucial
observation for hypersensitive response is that the necessary condition,
$\tau << k^{-1}$ or $\alpha << {(-lnf)}^{-1}$, can be satisfied by very small
forces for moderately small $\alpha$
because of the logarithmic dependence on $f$, see
Fig.~\ref{fig3}. In this
figure one also observes the region of ``normal" (i.e., linear)
response for higher forcing or frequency
and, more relevant to our preoccupation here, the
strongly nonlinear dependence at very low forcing. In fact in the limit
$f \rightarrow 0$,  one finds from  Eq.~(\ref{velocity})
that $\langle\dot{x}\rangle/v_0\approx [2\alpha \Gamma (ln f)^2]^{-1}$.
In other words, hypersensitive response is very pronounced in
this region, with the velocity picking up with an infinite derivative at $f=0$.

In conclusion, the procedure presented here has resolved all technical
problems related to steady state dichotomous dynamics, making possible
the analytic description of cases involving the crossing of fixed points
in the asymptotic regime.

This work was partially supported by the National Science Foundation
under grant Nos.  PHY-9970699  and DMS-0079478.

\end{document}